Classification:  Physical Sciences (Applied Physical Sciences)
Biological Sciences (Biophysics)

# Kinks, Rings, and Rackets in Filamentous Structures

Adam E. Cohen[1] and L. Mahadevan[2]

[1]Department of Physics, Stanford University, Stanford, CA 94305
[2]*To whom correspondence should be sent*:
Division of Engineering and Applied Sciences
Harvard University,
Cambridge, MA 02138
phone: (617) 496-9599
email: lm@deas.harvard.edu

Text: 14 pages
Figures: 4

Abstract: 61 words

| | |
|---|---|
| All text characters (including title page, abstract, legends, references) *plus* spaces | 32,468 |
| Fig. 1 (2-column, 20 cm high = 360 x 20) | 7,200 |
| Fig. 2 (1-column, 6 cm high = 180 x 6) | 1,080 |
| Fig. 3 (1-column, 6 cm high = 180 x 6) | 1,080 |
| Fig. 4 (1-column, 6 cm high = 180 x 6) | 1,080 |
| Equations [sixteen one 1-line eq., 1-col. = (16 x 60)] | 960 |
| **Space Allowance** | |
| 4 single-column figures (4 x 120) | 480 |
| 1 double-column figure (1 x 240) | 240 |
| 16 single-column equations (16 x 120) | 1920 |
| **Total characters in paper (must not exceed 47,000)** | 46,508 |

**Abstract**

Carbon nanotubes and biological filaments each spontaneously assemble into kinked helices, rings, and "tennis racket" shapes due to competition between elastic and interfacial effects. We show that the slender geometry is a more important determinant of the morphology than any molecular details. Our mesoscopic continuum theory is capable of quantifying observations of these structures, and is suggestive of their occurrence in other filamentous assemblies as well.

**I. Introduction**

Small self-assembled structures are common in biology, chemistry and condensed matter physics. The rich morphology that these structures exhibit arises from a combination of short and long range forces, often mediated by the presence of thermal fluctuations and hydrodynamic forces. From a geometrical perspective, the simplest self-assembling structures arise from the interaction between particles (monomers), and lead to the formation of globules and filaments. At the next level of complexity, filaments can aggregate into higher-order structures such as helices, rings, tapes, sheets, etc. At the mesoscopic level, the interactions within a filament may be represented by long wavelength elastic deformations due to stretching, bending and shear, while the complex interactions between filaments can be replaced by a simple short-range adhesive potential. In a variety of systems such as organic and inorganic nanotubes as well as stiff biopolymers, the stretching and shear deformation modes are energetically expensive relative to the bending modes, so that the filaments may be approximated as inextensible. In such cases, the competition between bending elasticity and adhesion is sufficient to explain the shapes seen in filamentous aggregates. We address the equilibrium morphologies of kinks, rings and rackets in these systems.

First we review the linear mechanics of thin rods and consider the conditions under which a classical description suffices. The stiffness of a rod is measured by its bending constant, $YI$, where $Y$ [N m$^{-2}$] is the Young's modulus of the material and $I$ [m$^4$] is the area moment of inertia, given by the second moment of the mass distribution in a cross-section perpendicular to the axis of symmetry.[1] The bending energy per unit length is $YI\kappa^2/2$, where $\kappa$ is the curvature.

Thermal fluctuations bend a rod on the scale of its *persistence length*, $l_p = YI/k_BT$. These are the approximate room temperature persistence lengths of the rods considered below: *Limulus* acrosome: 2.7 m; single walled carbon nanotube: 45 µm; S-hemoglobin fiber: 240 µm; microtubule: 6 mm. The persistence length of a multiwalled nanotube depends on the number of shells, but is always much greater than that of a single walled nanotube. In each case the persistence length is much greater than the actual length of the rod, so thermal fluctuations are negligible. Thermal fluctuations may be introduced as a weak perturbation in these systems,[2] but we do not do so here.

**II. Kinks**

We start by examining kinked helices in multiwalled carbon nanotubes (MWNTs) and in the acrosome of horseshoe crab (*Limulus*) sperm. MWNTs are fibers composed of concentric graphene tubules. They show promise as components of nanoelectronic devices, field-emission displays, and high-strength composites. MWNTs are usually fairly straight, but under some growth conditions tubes form with a corkscrew shape.[3,4,5] The tubes grow out of molten catalyst particles that have been supersaturated with carbon, and the corkscrew shape arises when there is a nonuniform rate of deposition of carbon around the circumference of the tube.[6] Close examination of a corkscrew MWNT often shows that the tube is composed of relatively straight sections joined at kinks.[7,8] An example of this is shown in Figure 1a.

The acrosomal process of a *Limulus* sperm is a ~50 µm-long rod of bundled actin filaments. In a free-swimming sperm the acrosome is coiled around the base of the sperm. When the sperm encounters an egg, a calcium signal causes the acrosome to uncoil so that it juts out the front of the sperm and harpoons the egg.[9] Interestingly, the coiled acrosome is also composed of straight sections joined at kinks, as shown in Figure 1b.[10]

The occurrence of a kinking instability in helices of these two seemingly dissimilar rods suggests that the kinks may arise through a common mechanism. Ihara and coworkers[11] and Dunlap[12,13] have proposed a model of helix-formation in carbon nanotubes based on pentagon-heptagon paired defects (PHPDs). Putting a pentagon and heptagon of carbon atoms on diametrically opposite sides of a nanotube introduces a kink into the nanotube. Arrays of such kinks form a helix. Unlike the mechanism discussed in this paper, forming PHPDs requires breaking covalent bonds. The PHPD mechanism also does not account for the observed periodicity of the kinks, nor why PHPDs should align in successive shells of a MWNT to produce localized kinks. Furthermore, it is not clear why kinked helices are observed in MWNTs but not in single walled nanotubes (SWNTs). Finally, the PHPD mechanism is specific to carbon nanotubes, and a different mechanism would be needed for the acrosome.

We propose a general model of kinking in fibrillar aggregates. Consider an aggregate of fairly inextensible fibers that are weakly coupled to each other. Both the concentric graphene shells in MWNTs and the actin filaments in the acrosome fit this description. Each fiber has corrugations along its length because it is composed of discrete molecular or atomic monomers; these corrugations reflect the periodic nature of the fiber and its interaction with its neighbors. Adjacent fibers are most stable when their corrugations are in registry, but this cannot occur everywhere along a bent aggregate. Bending or twisting introduces an *effective lattice mismatch* between fibers on the inside of the curve and those on the outside of the curve. Kinks develop where outer fibers slip one lattice constant behind their inner neighbors. The energy cost of introducing a kink is less than the energy gained by straightening segments between kinks.

Simple geometry determines the angle each kink subtends. Let $D$ be the distance between adjacent fibers, projected onto the plane of the curve, and $S$ be the period of the corrugations along a fiber. Bending the aggregate through an angle $\theta_k = S/D$ leaves each fiber exactly one corrugation behind its inner neighbor. The interaction energy *per unit length* between adjacent fibers, $U_{int}(l)$, is a periodic function of their relative axial displacement, with period $S$. We approximate this interaction with a simple sinusoidal potential:

$$U_{int}(l) = -\frac{\Delta\gamma}{2}\cos\left(\frac{2\pi D}{S}\theta(l)\right), \qquad [1]$$

where $\Delta\gamma$ measures the strength of the corrugations in the interaction potential and the contour of the aggregate is characterized by the angle, $\theta(l)$, between its orientation at position $l$ and the orientation of one end. We take $\theta(0) = 0$ so that the corrugations are in registry at the beginning of the aggregate. Each fiber experiences a bending energy as well as the interfacial energy, so its total energy *per unit length* is:

$$U(l) = \frac{YI}{2}\theta'(l)^2 - \frac{\Delta\gamma}{2}\cos\left(\frac{2\pi D}{S}\theta(l)\right), \qquad [2]$$

where the prime indicates a derivative with respect to length and $YI$ is the bending constant of a single fiber. To find the function $\theta(l)$ that minimizes $H = \int_0^L U(l)dl$, where $L$ is the total length of the aggregate, we set the variational derivative $\delta[H] = 0$. The equation of equilibrium for the aggregate is

$$YI\theta''(l) = -\frac{\chi}{2}\Delta\gamma\sin(\chi\theta(l)), \qquad [3]$$

where the new variable $\chi \equiv 2\pi D/S$. This equation of motion may be recognized as the Frenkel-Kontorova model.[14] The Frenkel-Kontorova model provides a nonlinear microscopic description of periodic dislocations that occur in lattice-mismatched epitaxial layers. Here the lattice-mismatch is replaced by a curvature-induced effective lattice mismatch. Srolovitz, Safran and Tenne[15,16] used the concept of effective lattice mismatch to develop a mesoscopic continuum model of kinking in thin 2-dimensional films, but they did not consider the mechanics in the vicinity of the kink as we do here.

When $\chi = 1$ the solutions of [3] correspond to the standard elasticae of a homogeneous isotropic rod (or equivalently to solutions of the simple pendulum, with $l$ being a time-like variable). In general $\chi \neq 1$ and a range of other interesting shapes results. For $\chi > 1$ we find *kinked aggregates*, where $\chi$ gives the number of kinks per loop of the aggregate. With the initial value $\theta(0) = 0$, the aggregate switches from being essentially straight, with small sinusoidal perturbations, to kinked at $\theta'(0) \geq (2\Delta\gamma/YI)^{1/2}$, where the equality corresponds to the separatrix solution of the elastica with a single loop (or equivalently, the solution for the pendulum that delineates the oscillatory solutions from the rotating solutions). As $\theta'(0)$ increases beyond $(2\Delta\gamma/YI)^{1/2}$ the aggregate adopts an ever more circular aspect.

When the kinks are far apart so that the sections between kinks are approximately straight, it is possible to solve analytically for the shape of a kink. Multiplying both sides of [3] by $\theta'(l)$ and integrating with the boundary conditions $\theta'(-\infty) = 0 = \theta'(\infty)$ yields

$$\theta(l) = \frac{4}{\chi}\tan^{-1}\left[\exp\left(l\chi\sqrt{\frac{\Delta\gamma}{2YI}}\right)\right]. \qquad [4]$$

Equation [4] shows that the kink occurs over a length $l_{kink} \approx \frac{1}{\chi}\sqrt{\frac{2YI}{\Delta\gamma}}$, as could be expected on dimensional grounds. The energy of a single kink is obtained by substituting the solution [4] into the energy functional [2] to yield

$$U_k = \frac{4}{\chi}\sqrt{2YI\Delta\gamma}. \qquad [5]$$

Figure 2 shows the curve obtained for $\chi = 8$, and $\theta'(0) = 1.0001 \times (2\Delta\gamma/YI)^{1/2}$. In real systems the corrugation potential is not a perfect sinusoid, so kinks will in general have a shape slightly different from that described by [4].

Structural data obtained by electron-microscopy allows us to apply this model to the *Limulus* acrosome.[17] The crosslinks between fibers have a period of $S = 55$ Å along a filament, and the separation between filaments is $D = 147$ Å. The ratio $S/D$ gives a kink-angle of $\theta_k = 0.37$ rad, or 21°, in reasonable agreement with the observed kink angle of 24°.

We can also estimate the distance between kinks from the molecular structure. A cross-section of the acrosome shows that the actin fibers are hexagonally packed. For all kinks to lie in the same plane, the acrosome must twist through a multiple of 60° between kinks. The actin monomers are spaced by 27 Å, and the crosslinking protein *scruin* introduces a twist of 0.23° per monomer when the acrosome is coiled.[18] Thus, the spacing between kinks is roughly 27 Å × 60° / 0.23° = 7000 Å. From the kink-angle and the spacing between kinks we find that the coiled acrosome makes one loop every 10 μm of its length, to produce a coil with a diameter of 3.2 μm. This coil just fits inside the head of the sperm. Thus the molecular dimensions of the acrosomal constituents interact to set the size of the entire coiled acrosome.

The modified Frenkel-Kontorova model is more difficult to apply quantitatively to MWNTs because the constituent fibers (SWNTs) are concentric rather than adjacent. Nonetheless, similar reasoning applies. Thin graphite sheets form a kinked twin matrix boundary of 20°48' about $[1\bar{1}00]$,[19] which is very close to the kink-angle observed in MWNT helices. The surfaces of a MWNT on the inside and outside of a curve develop these kinks to relax strain. It is noteworthy that in cross-section MWNTs also often appear polygonal rather than circular. This polygonalization cannot be explained in terms of pentagon-heptagon defects, but arises naturally in a model based on curvature-induced lattice mismatch. The twin matrix boundary angle of 20°48' implies that a cross-section of a MWNT should have roughly 18 edges. In practice some of these edges are typically too short to observe. Polygonalization also occurs in nested fullerenes[20,21] and WS$_2$ nanoparticles.[22] TEM pictures of MWNTs show that there is also some delamination of the graphene sheets and buckling in the popliteal region of each kink. These effects occur because of topological constraints on the graphene sheets in MWNTs, and are better explained in terms of buckling of a hollow tube.

Although we have focused on lattice-slip in the presence of spontaneous curvature as the source of mechanical nonlinearity leading to planar kinks, the same mechanisms will give rise to kinks in nonplanar fibrillar aggregates because of the competition between bend, twist and adhesion. More generally these localized structures arise in aggregates because of the presence of a non-convex bending energy functional or equivalently, by virtue of simple dimensional arguments that penalize kinks and bends differently.

In MWNTs and acrosomes the growth conditions impose a mean curvature; the array of kinks minimizes the energy while maintaining this curvature. At finite temperature thermal fluctuations may also nucleate kinks. The density of thermally activated kinks is proportional to $e^{-U_k/k_BT}$, where $U_k$ is given by [5]. For the present systems $U_k >> k_BT$, so thermally activated kinks may be neglected. When $U_k \sim k_BT$, the density of kinks can be calculated using the methods developed by Büttiker and Landauer for overdamped sine-Gordon solitons.[23]

**III. Rings**

Another process determined by competition between interfacial and elastic effects is the formation of rings from microscopic rods. An ultrasonically induced cavitation bubble collapses around a rope of single-walled carbon nanotubes (SWNTs), causing the rope to form a ring (Fig 1c).[24,25] Tubulin molecules polymerize inside red blood cells of birds and reptiles until they encounter the cell membrane, whereupon the tubulin loops into a ring around the equator of the cell (Fig 1d).[26] When fibers of defective S-hemoglobin encounter the membrane of an erythrocyte, the fibers remain straight and deform the membrane into the shape characteristic of sickle-cell anemia. In each case, the length of a nanoscale rod exceeds the diameter of a bounding surface, so that the interface applies a compressive load on the nanoscale rod. What determines whether the rod remains straight (as in sickle cell disease) or buckles (as in microtubules and SWNTs)? And if the rod buckles, does it eventually form a ring, or does it snap back to its straight state?

The system is parameterized by the ratio of the contour length of the rod, $l$, to the diameter of the bounding surface, $d$. An increasing value of $l/d$ applies equally to a rod of fixed length in a shrinking boundary (e.g. a SWNT inside of a cavitation bubble) and a growing rod in a fixed boundary (e.g. microtubules in vesicles, S-hemoglobin in erythrocytes). We distinguish three cases. 1) If the rod is stiff, the rod remains straight and together with the bounding surface forms a shape resembling the Greek letter ϕ for all $l/d > 1$. 2) If the rod is slightly less stiff, interfacial tension causes the rod to buckle at $l/d = 1$. As $l/d$ increases, strain builds up in the rod until its ends poke through the interface. The rod then snaps back to form a ϕ shape. 3) If the rod is sufficiently flexible, interfacial tension forces the rod to bend all the way around into a loop.

The occurrence of the initial buckling transition is determined by a competition between bending induced by buckling of the rod and the maximum compressive force that the interface can exert. The Euler buckling load for a simply supported rod of length $l$ is[27]

$$F_b = \frac{YI\pi^2}{l^2}.$$

Odijk has calculated a correction to this expression of order $l/l_p$.[2] We use the classical result because, for the systems considered here, $l \ll l_p$. The maximum force from the interface depends on whether the interface is a simple liquid or a biological membrane.

### IIIa. Rods in liquid drops

A liquid interface exerts a compression force $F_c = 2\pi r \gamma \cos\theta$, where $r$ is the radius of the rod, $\gamma$ is the tension of the interface and $\theta$ is the contact angle at the interface, assumed to be its equilibrium value. If $F_c > F_b$, the rod buckles; otherwise it remains straight. Balancing the two forces yields a critical length

$$l_c = \left(\frac{\pi YI}{2r\gamma\cos\theta}\right)^{1/2}, \qquad [6]$$

below which the rod remains straight and above which the rod buckles. For a typical SWNT in a cavitation bubble in water, $YI \sim 1.9 \times 10^{-25}$ N m²,[28,29,30] $r \sim 0.7$ nm, $\gamma \sim 70$ mN/m and $\cos\theta \sim 1$, so $l_c \sim 155$ nm. In the Appendix we calculate the energy and stability of an elastic rod confined by a spherical interface of prescribed surface energy, to determine if, after the SWNTs buckle, they form a ring at $l/d = \pi$, or if they puncture the interface at some $l/d$ between 1 and $\pi$.

Figure 3 shows the computed stability diagram for an elastic rod in a droplet. In the region marked $\alpha$ both the buckled conformation and the $\phi$ shape are stable. The conformation is determined by the direction from which the system enters the region $\alpha$.

A growing rod in a droplet of fixed size traverses the diagram along a line starting from the origin with a slope given by $d/l_c$. A rod of fixed length inside a shrinking droplet (or bubble) traverses the diagram along a horizontal line from left to right. We see that the rod buckles if $l > l_c$, but will only form a loop provided that $l > l_c\sqrt{2}$. Martel *et al.* measured a rope diameter of 30 nm in their rings of SWNTs. Assuming a SWNT diameter of 1.4 nm and that the SWNTs in the rope *can* slide relative to each other (so $l_c \sim n^{1/4}$, see below), we find that the critical rope-length for loop formation in water is 1 μm, or that the minimum ring radius should be 160 nm. Martel *et al.* found that no rings of SWNTs with radii less than 250 nm, in qualitative agreement with our model. The discrepancy between the result of Martel *et al.* and our prediction may be due to some inter-tube shear, nonzero contact angle between the water and the rope, and a reduction of the surface tension of the water from dissolved $H_2SO_4$. Thermal fluctuations are insufficient to wrap the rod into a loop, contrary to the model of thermally activated ring-formation in SWNTs proposed by Sano and coworkers.[31]

**IIIb. Rods in vesicles and cells**

We now turn to the case of rods confined to vesicles and cells that have a membranous outer layer. Bilayer membranes have a small but finite bending stiffness in addition to a surface tension. The bending stiffness distributes a localized force over an area of radius $r_{eff} = (2\kappa/\mu)^{1/2}$, where $\kappa$ and $\mu$ are the bending modulus and tension, respectively, of the membrane.[32] Provided that the radius of the rod is much less than $r_{eff}$, the maximum compression force the membrane can exert is $F_c = 2\pi r_{eff} \mu$, independent of the radius of the rod. Substituting this force into the expression for the Euler buckling load yields

$$l_c = \left(\frac{\pi}{2} YI\right)^{1/2} (2\kappa\mu)^{-1/4}. \qquad [7]$$

If the rod is shorter than the critical length, then the membrane extends a sheath around its points of contact with the rod to form a ϕ shape. This case is equivalent to the rod puncturing the surface of a liquid interface.

It is known that tubulin polymerizes into an equatorial microtubule ring during the morphogenesis of avian erythrocytes (similar to that shown in Fig. 1d), and is responsible for the initiation of the bi-concave shape that is crucial for the cell to be able to navigate through narrow capillaries by deforming easily. Ironically, in sickle cell erythrocytes, S-hemoglobin forms fibrillar aggregates that change the shape of the cell so much that it is stiffened to the point that it can no longer negotiate the capillary vessels. A single S-hemoglobin fiber ($YI \sim 1 \times 10^{-24}$ N m$^2$, r $\sim 10.5$ nm)[33], of length equal to the diameter of a human erythrocyte ($l \sim 7.5$ μm), buckles under a compressive force of $F_c \sim 0.175$ pN. Yet by our estimate a red cell membrane ($\kappa \sim 2 \times 10^{-19}$ N m,[34] $\mu \sim 2 \times 10^{-6}$ N/m) can sustain a force of $F_c \sim 6$ pN (optical tweezers experiments measure $F_c \sim 20$ pN[35]) so it is likely that sickle cell erythrocytes deform only because S-hemoglobin fibers aggregate into ropes which are much stiffer than a single fiber.

To determine the scaling of the critical length for buckling with the number of fibers $n$ in a rope, we consider the extreme cases of (a) fibers which are free to slide past each other so that $F_b \propto n$ and (b) fibers that are tightly crosslinked so that $F_b \propto n^2$. The force of surface tension is proportional to the circumference of the rope and hence for a liquid $F_c \propto n^{1/2}$, while for a membrane $F_c$ is independent of $n$. Equation [6] shows that the critical length of a rope bounded by a liquid interface scales as $l_c \propto n^{1/4}$ if the fibers can slide past each other, or $l_c \propto n^{3/4}$ if the fibers are rigidly crosslinked. Equation [7] shows that the critical length $l_c$ of a rope bounded by a membrane scales as $l_c \propto n^{1/2}$ if the fibers can slide past each other, or $l_c \propto n$ if the fibers are rigidly crosslinked.

This scaling may be important in sickle-cell disease. Since an erythrocyte membrane can resist a point-force roughly 100 times larger than the buckling force of a single S-hemoglobin fiber, if the fibers in an aggregate can slide past each other it would take roughly 100 fibers to sickle a cell. However, if the fibers in an aggregate are crosslinked, then it would take only 10 fibers to sickle a cell. Therapeutic agents that allow fibers to slide relative to each other may decrease sickling and thus decrease sickness.

Although there is evidence confirming the relation between membrane tension and the buckling of microtubules,[36] no equivalent experiments have been performed for SWNTs. If the buckling mechanism of ring-formation is correct, then rings of carbon nanotubes should also form in vesicles and in fluid-fluid colloidal dispersions in which the colloidal phase wets the nanotubes better than does the bulk phase. Tuning the wettabilities of the two fluids should provide control over the diameter of the rings. This may be a route to colloidal particles with unusual morphologies.

**IV. Rackets**

A third shape observed in both carbon and other nanotubes and biological microtubules is the "tennis racket,"[37] which occurs when the rod folds into a closed figure with both its ends pointing in the same direction (Fig. 1d, e). Recent work that we became aware of after completing our own focuses on aspects of this shape while considering the collapse of semi-flexible polymer chains.[38,39] Here we will consider only the zero-temperature aspects of this problem. Rod-rod attraction seeks to zip up the loop, while elastic bending resists this tendency; a balance between these two effects determines the size of the loop. The interfacial energy is $E_{int} = -l_{int} \gamma_{adh}$, where $l_{int}$ is the length of the rod-rod contact line and $\gamma_{adh}$ measures the rod-rod interaction energy. The elastic energy in the loop scales as $E_{el} \sim YI/l_{loop}$, where $l_{loop}$ is the circumference of the loop. Noting that $l_{int} + l_{loop} = constant$, we can estimate the size of the loop by minimizing $E_{el} + E_{int}$ with respect to $l_{loop}$. This yields

$$l \sim \left( \frac{YI}{\gamma_{adh}} \right)^{1/2} \quad [8]$$

where $l$ is a characteristic dimension of the loop. Finite-element simulations confirm that the width $w$ of the loop, measured perpendicular to the axis of symmetry, is $w \sim \sqrt{YI/\gamma_{adh}}$ and its length, $l$, measured from the point where the ends of the rod meet to the top of the loop is $l \sim 2\sqrt{YI/\gamma_{adh}}$.

The rod-rod adhesion energy for a SWNT with a radius of 0.7 nm is $\gamma_{adh} \sim 2.8$ eV/nm,[46] giving $w \sim 20$ nm. The carbon nanotube "tennis racket" in figure 1d has $w = 60$ nm, which is in reasonable agreement with theory, given the uncertainties in the mechanical constants and that the nanotube in question seems to be a rope of several nanotubes along part of its length.

By measuring the diameter of the microtubule loop in figure 1d ($w \sim 1.75$ μm) and using the known stiffness of microtubules ($YI \sim 2.6 \times 10^{-23}$ N m$^2$),[40] we estimate the interaction energy of two microtubules as 60 meV/nm, or 2.3 $k_b$T/nm. This measurement can be compared with a simple estimate of the van der Waals attraction between two protein rods in solution. The interaction energy per unit length of two parallel rods is[41]

$$\gamma_{adh} = \frac{H_A}{12\sqrt{2}\delta^{3/2}}\left(\frac{r_1 r_2}{r_1 + r_2}\right)^{1/2} \equiv U\left(\frac{2 r_1 r_2}{r_1 + r_2}\right)^{1/2}, \qquad [9]$$

where $H_A$ is the Hamaker constant, $r_1$ and $r_2$ are the radii of the rods, and $\delta$ is the distance of closest approach between the rods. We ignore double-layer and hydration forces. Calculations give $H_A = 3.1$ $k_b$T for proteins interacting through water, while attempts to fit experimental data give $H_A = 1$-$10$ $k_b$T.[42] The value of $\delta$ is also uncertain, ranging from 0.1 nm to 0.3 nm, and varying with the ionic strength of the solution. We choose $H_A = 3$ $k_b$T, $\delta = 0.3$ nm, and $r_1 = r_2 = 12.5$ nm, which yields $\gamma_{adh} = 70$ meV/nm. Thus the two methods of estimating rod-rod attraction (analysis of tennis racket shapes and van der Waals energy) yield similar results.

To understand the dependence of the size of the racket head on the radius of the rod, we observe that the adhesion energy depends on the rod-radii because thicker rods have more atoms in proximity to the interface than do thinner rods. For the case of a rod of constant radius doubled over onto itself, equation [9] gives $\gamma_{adh} \approx U r^{1/2}$. The stiffness of hollow tubes is $YI = Y\pi r^3 t$, where $t$ is the wall-thickness, and the stiffness of solid rods is $YI = Y\pi r^4/4$. Inserting the scaling laws for adhesion and stiffness into equation [8] yields $w \propto r^{5/4}$ for tubes and $w \propto r^{7/4}$ for solid cylinders.

A racket made of a solid rod bends smoothly just like a tennis racket. However, a racket made from a hollow tube whose radius exceeds a critical value is unstable to the formation of a kink at the apex of the racket just like a strongly bent drinking straw. Brazier showed that a tube develops a kink when its curvature exceeds a critical value[43]

$$\kappa_c = \frac{1}{3r^2}\sqrt{\frac{2t}{(1-\sigma^2)}}, \qquad [10]$$

where $t$ is the thickness of the tube wall (assumed to be much less than $r$) and $\sigma$ is its Poisson ratio. Iijima and coworkers observed[44] and Yakobson and coworkers simulated[45] this phenomenon in SWNTs and found that

$$\kappa_c = \frac{0.0388 nm}{r^2}. \qquad [11]$$

Thus, since the radius of curvature of the racket head grows as $r^{5/4}$ (see above), while the minimum radius of curvature that avoids kinking grows as $r^2$, we can combine [8] and [10] to show that the tennis racket shape is stable against kink formation at its apex if

$$r < \left(\frac{t}{6}\right)^{4/3}\left(\frac{2\pi Y}{U(1-\sigma^2)}\right)^{2/3}. \qquad [12]$$

Using [11] for the critical curvature of a SWNT and the value $U = 3.35$ eV nm$^{-3/2}$,[46] yields a maximum radius for smoothly deformed SWNT of 0.83 nm. This radius is well within the range of accessible SWNT radii, and thus buckled tennis racket shapes should be observable in AFM scans of carbon nanotubes.

## V. Conclusions

We have shown that similar features occur in carbon nanotubes and cellular organelles. These features result from simple coarse-grained mechanical properties of nanoscale rods, and are independent of the molecular details of the media in which they occur, but are dominated by the geometry and can be explained with simple mechanical models. Kinked helices, rings, and tennis racket shapes may also appear in other nanoscale rods which are in us (as assemblies of biomacromolecules such as actin bundles, microtubules etc.), and around us, occurring naturally (montmorillonite clays and vanadium pentoxide nanowires), and in laboratories (semiconductor nanowires, and molecular dye-aggregates).

Our analysis has focused on the most common equilibrium filamentous aggregate-structures that have been observed, and can be used either directly to predict the quantitative aspects of the morphology, or inversely to determine the mechanical properties of the filaments and their interactions. An obvious next step is to address the kinetics of formation of these aggregates; however this is a much more difficult problem since the mechanism by which the rings, rackets and kinks are formed is crucially affected by the detailed temporal sequence of events, as evidenced in the phase diagram shown in Figure 4.

**Acknowledgments**: AEC gratefully acknowledges support from the Marshall Fellowship and thanks members of the Bioproducts and Nanotechnology Group at Cambridge for interesting discussions. LM acknowledges the support of the US Office of Naval Research through a Young Investigator Award, the US National Institutes of Health, and the Schlumberger Chair Fund.

**Appendix – Energy of an elastic rod confined to a liquid droplet**

Here we calculate the elastic energy and stability of a rod confined to a liquid droplet. A similar line of reasoning can be applied to a rod inside a vesicle or a rod in a cavity in a liquid that does not wet the rod.

There are two distinct regimes we must consider: 1) For $1 < l/d < \pi/2$ the only contact is between the ends of the rod and the poles of the drop. 2) For $\pi/2 < l/d < \pi$ the entire length of the rod contacts the boundary of the drop along an arc of a great circle. We now calculate the energy of the buckled rod in these two regimes and examine the stability of the configuration to puncturing.

*Regime 1: $1 < l/d < \pi/2$*. Unlike a classical spring, a buckled rod exerts a restoring force, $F_b$, that is to lowest order independent of the deformation distance, $z$:

$$F_b = \frac{YI\pi^2}{l^2} + \frac{YI\pi^2}{2l^2}\left(\frac{z}{l}\right). \qquad [A1]$$

This last expression is a direct consequence of the nature of the instability of a buckled rod that takes the form of a supercritical pitchfork bifurcation.[47]

To find the elastic energy in the rod we calculate the work done in moving one endpoint from $z = l$ to $z = d$ while keeping the other endpoint fixed at $z = 0$:

$$E_{el} = \int_0^{l-d} F_b(z)dz \approx \frac{YI\pi^2}{l^2}(l-d) + \frac{YI\pi^2}{4l^3}(l-d)^2. \qquad [A2]$$

*Regime 2: $\pi/2 < l/d < \pi$*. The energy density of a bent rod is $\frac{YI}{2}\kappa^2$, where $\kappa$ is the curvature. Assuming that the rod is an arc of a great circle with constant radius of curvature $\kappa^{-1} = d/2$ we get:

$$E_{el} = \frac{2YIl}{d^2}. \qquad [A3]$$

Figure A1 shows the elastic energy of a rod confined to a sphere, both from analytical predictions [A1] and [A2], and from finite-element simulations run using the "Surface Evolver" package.[48,49]

Interfacial energy acts against the elastic energy to keep the rod in the drop. The interfacial energy is, $E_{int} = -2\pi rl\gamma\cos\theta$, where $l$ is now the length of rod inside the drop. The total energy of the system is $E = E_{el} + E_{int}$ and computing it for different configurations allows us to construct the phase diagram depicted in Figure 3. If $\partial E / \partial l$ is positive, then the system can lower its energy by expelling some of the rod from the drop since those parts of the rod outside the drop do not contribute to the elastic and interfacial energy; this scenario is apt for very stiff rods. However very flexible rods can easily accommodate the curvature induced by the liquid interface, and so remain completely embedded in the drop.

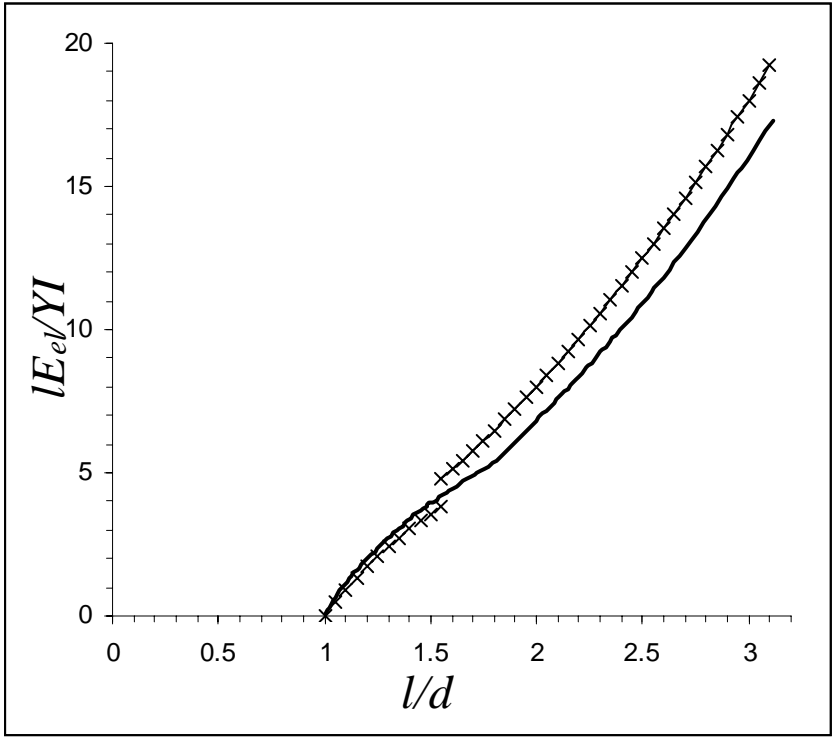

**Figure A1**: Dimensionless elastic energy, $lE_{el}/YI$, ($E_{el}$ is the energy), in an elastic rod confined to a sphere as a function of the ratio of rod-length, $l$, to sphere diameter, $d$. Crossed lines: analytical predictions [A2] and [A3]; thick line: numerical simulation. The analytical prediction is accurate for $1 < l/d < \pi/2$, but overestimates the energy for $\pi/2 < l/d < \pi$ because the ends of the rod cut cords through the inside of the drop, thus lowering the elastic energy below that of a perfectly circular arc.

**Figure 1**

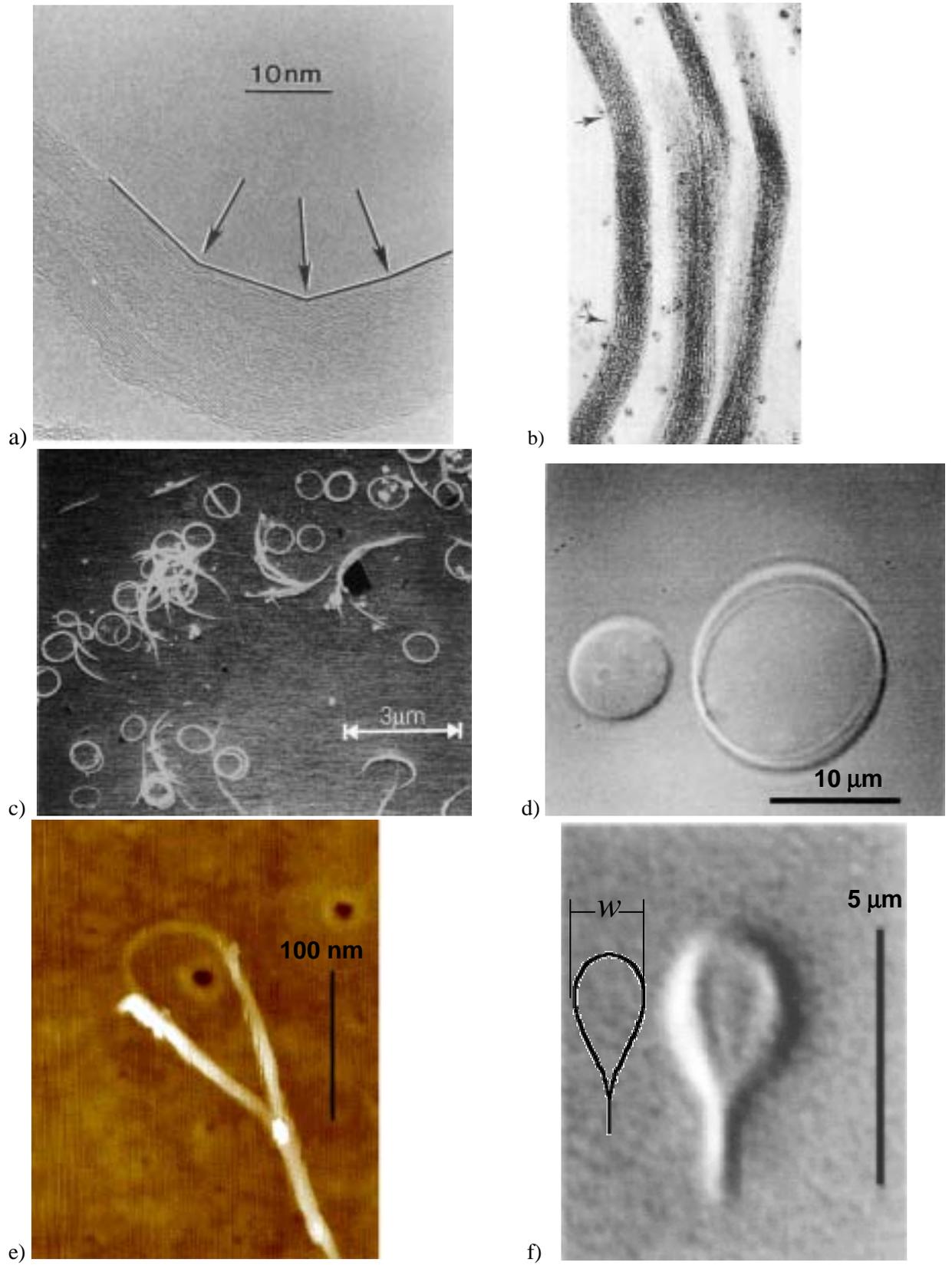

**Figure 1**: Analogous structures in carbon nanotubes (left) and cellular organelles (right).
a) Kinks in a helical MWNT.[8]
b) Kinks in the acrosomal process of a *Limulus* sperm.[18]
c) Rings of SWNTs formed when cavitation bubbles collapsed around ropes of SWNTs causing the ropes to buckle.[24]
d) Ring of tubulin formed when a tubulin rod grew to a length exceeding the diameter of a bounding lipid vesicle and the rod buckled.[36]
e) SWNT "tennis racket" observed in a sample of HiPCo SWNTs[50] after 30 min. sonication in dichloroethane.
f) Tubulin "tennis racket" observed in a tubulin rod that had buckled inside of a vesicle.[51]

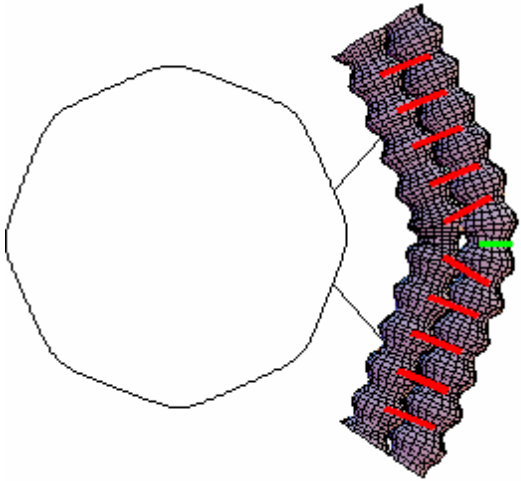

**Figure 2**: Energy-minimizing shape of a bent aggregate of corrugated fibers. The corrugation represents the periodicity of the inter-filament interactions. The fibers cannot maintain registry around a curve. Forming a kink minimizes the length that is out of registry but introduces an elastic penalty from the high curvature. Shown above is a kinked structure corresponding to the solution of Equation [5] with χ=8, $\theta'(0) = 1.0001 \times (\frac{2\Delta\gamma}{YI})^{1/2}$. The outer fiber has one additional corrugation at the kink, indicated by the green bar.

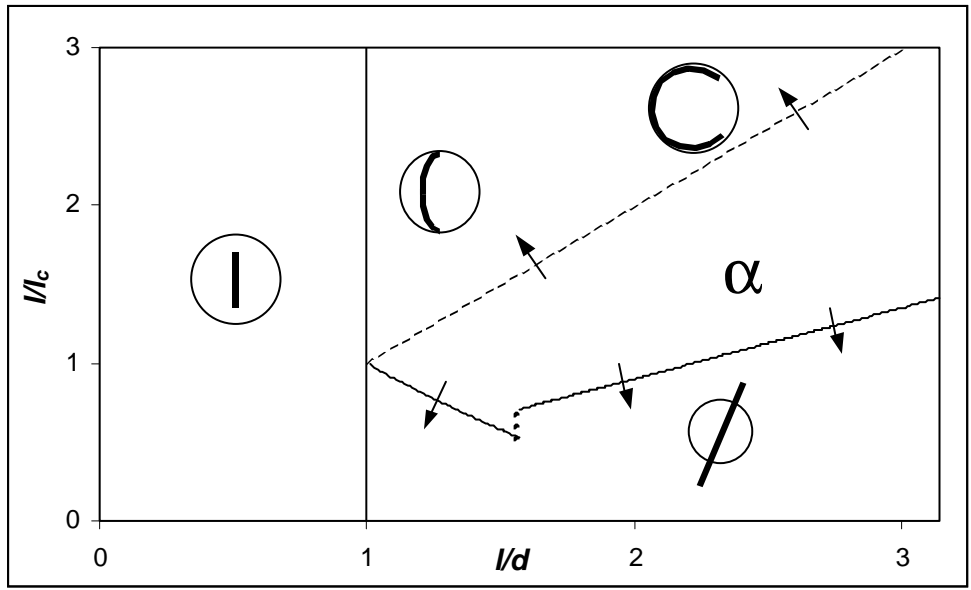

**Figure 3**: Stability diagram of a rod bounded by a liquid interface. The critical length, $l_c$ is defined in [6], $d$ is the diameter of the droplet, and $l$ is the length of the rod. There is an intermediate region, $\alpha$, in which both the curved-rod and the straight-rod conformations are stable. The boundaries are calculated analytically as described in the Appendix. The discontinuity at $l/d = \pi/2$ arises from different approximations required when the rod contacts the drop only at its poles verses when the rod contacts the drop along an arc of a great circle.

[1] For molecular-scale objects such as carbon nanotubes, $Y$ and $I$ are not separately well-defined, but their product is.
[2] T. Odijk, "Microfibrillar buckling within fibers under compression," *J. Chem. Phys.* **108**, 6923 (1998).
[4] R. Gao, Z. L. Wang, and S. Fan, "Kinetically controlled growth of helical and zigzag shapes of carbon nanotubes," *J. Phys. Chem. B* **104**, 1227 (2000).
[4] K. Hernadi, L. Thiên-Nga, and L. Forró, "Growth and microstructure of catalytically produced coiled carbon nanotubes," *J. Phys. Chem. B* **105**, 12464 (2001).
[5] M. Zhang, Y. Nakayama, and L. Pan, "Synthesis of carbon tubule nanocoils in high yield using iron-coated indium tin oxide as catalyst," *Jpn. J. Appl. Phys.* **39** L1242 (2000).
[6] S. Amelinckx et al. "A formation mechanism for catalytically grown helix-shaped graphite nanotubes," *Science* **265**, 635 (1994).
[7] X. B. Zhang et al., "The texture of catalytically grown coil-shaped carbon nanotubules," *Europhys. Lett.* **27**, 141 (1994).
[8] D. Bernaerts et al., "Electron-microscopy study of coiled carbon tubules," *Phil. Mag. A* **71**, 605 (1995).
[9] L. G. Tilney, "Actin filaments in the acrosomal reaction of *Limulus* sperm," *J. Cell. Biol.* **64**, 289 (1975).
[10] D. J. Derosier and L. G. Tilney, "How to build a bend into an actin bundle," *J. Mol. Biol.* 175, 57 (1984).
[11] S. Ihara, S. Itoh, and J.-I. Kitakami, "Helically coiled cage forms of graphitic carbon," *Phys. Rev. B* **48**, 5643 (1993).
[12] B. I. Dunlap, "Relating carbon tubules," *Phys. Rev. B* **49**, 5643 (1994).
[13] B. I. Dunlap, "Constraints on small graphitic helices," *Phys. Rev. B* **50**, 8134 (1994).
[14] Frenkel, T. Kontorova, *Phys. Z. Sowjet.* **13**, 1 (1938). F.C. Frank, J.H. van der Merwe, *Proc. R. Soc. (A)* **198**, 205 (1949).
[15] D. J. Srolovitz, S. A. Safran, and R. Tenne, "Elastic equilibrium of curved thin films," *Phys. Rev. E* **49**, 5260 (1994).
[16] S. A. Safran, "Curvature elasticity of thin films," *Advances in Physics* **48**, 395 (1999).
[17] M. B. Sherman et al. "The three-dimensional structure of the *Limulus* acrosomal process: a dynamic actin bundle," *J. Mol. Biol.* **294**, 139 (1999).

[18] D. DeRosier, L. Tilney, and P. Flicker, "A change in the twist of the actin-containing filaments occurs during the extension of the acrosomal process in *Limulus* sperm," *J. Mol. Biol.* **137**, 375 (1980).

[19] B. T. Kelly, The Physics of Graphite, (Applied Science Publishers: London, 1981), p. 37.

[20] D. J. Srolovitz, S. A. Safran, M. Homyonfer, and R. Tenne, "Morphology of nested fullerenes," *Phys. Rev. Lett.* **74**, 1779 (1995).

[21] V. P. Dravid *et al.*, "Buckytubes and derivatives: their growth and implications for buckyball formation," *Science* **259**, 1601 (1993).

[22] U. S. Schwarz, S. Komura, and S. A. Safran, "Deformation and tribology of multi-walled hollow nanoparticles," *Europhys. Lett.* **50**, 762 (2000).

[23] M. Büttiker and R. Landauer, "Nucleation theory of overdamped soliton motion," *Phys. Rev.* A 23, 1397 (1981).

[24] R. Martel, H. R. Shea, and P. Avouris, "Rings of single-walled carbon nanotubes," *Nature* **398**, 299 (1999).

[25] R. Martel, H. R. Shea, and P. Avouris, "Ring formation in single-wall carbon nanotubes," *J. Phys. Chem. B* **103**, 7551 (1999).

[26] B. Winckler and F. Solomon, "A role for microtubule bundles in the morphogenesis of chicken erythrocytes," *Proc. Natl. Acad. Sci. USA* **88**, 6033 (1991).

[27] Landau, L. D. and Lifshitz, E. M (1970) *Theory of Elasticity*, transl. Sykes, J. B. and Reid, W. H. (Pergamon Press, New York).

[28] J.-P. Salvetat, G.A.D. Briggs, J.-M. Bonard, R.R. Bacsa, A.J. Kulik, T. Stöckli, N.A. Burnaham, and L. Forró, "Elastic and shear moduli of single-walled carbon nanotube ropes," *Phys. Rev. Lett.* **82** 944-947 (1999)

[29] J.-P. Salvetat, J.-M. Bonard, N.H. Thomson, J.J. Kulik, L. Forró, W. Benoit, and L. Zuppiroli, "Mechanical properties of carbon nanotubes," *Appl. Phys. A* **69** 255-260 (1999).

[30] S. Akita, H. Nishijima, T. Kishida, and Y. Nakayama, "Influence of force acting on side face of carbon nanotube in atomic force microscopy," *Jpn. J. Appl. Phys.* **39** 3724-3727 (2000).

[31] M. Sano, A. Kamino, J. Okamura, and S. Shinkai, "Ring closure of carbon nanotubes," *Science* **293**, 1299-1301 (2001).

[32] E. Evans and A. Yeung, "Hidden dynamics in rapid changes of bilayer shape", Chem. Phys. Lipids, **73**, 39-56 (1994).

[33] J. Wang et al., "Micromechanics of isolated sickle cell hemoglobin fibers: bending moduli and persistence lengths," *J. Mol. Biol.* **315**, 601 (2002).

[34] J. Sleep, D. Wilson, R. Simmons, and W. Gratzer, "Elasticity of the red cell membrane and its relation to hemolytic disorders: an optical tweezers study," *Biophys. J.* **77**, 3085 (1999).

[35] S. Henon, G. Lenormand, A. Richer, and F. Gallet, "A new determination of the shear modulus of the human erythrocyte membrane using optical tweezers," *Biophys. J.* **76**, 1145 (1999).

[36] D. Kuchnir Fygenson, M. Elbaum, B. Shraiman, and A. Libchaber, "Microtubules and vesicles under controlled tension," *Phys. Rev. E* **55**, 850 (1997).

[37] In old-fashioned tennis rackets, the frame is made of a piece of wood bent over onto itself. The wood minimizes its elastic energy by adopting the same shape as occurs in nanotubes and microtubules.

[38] B. Schnurr, F. C. MacKintosh, and D. R. M. Williams, "Dynamical intermediates in the collapse of semiflexible polymers in poor solvents," *Europhys. Lett.* **51**, 279 (2000).

[39] B. Schnurr, F. Gittes, and F. C. MacKintosh, "Metastable intermediates in the condensation of semiflexible polymers," arXiv:cond-mat/0112288 v2 17 Mar 2002.

[40] M. Elbaum, D. K. Fygenson, and A. Libchaber, "Buckling Microtubules in Vesicles," *Phys. Rev. Lett.* **76**, 4078 (1996).

[41] J. N. Israelachvili, Intermolecular and Surface Forces (Academic Press, New York), 1992, p. 177).

[42] A. Tardieu *et al.*, "Proteins in solution: from X-ray scattering intensities to interaction potentials," *J. Crystal Growth* **196**, 193 (1999); A. J. Rowe, "Probing hydration and the stability of protein solutions—a colloid science approach," *Biophys. Chem.* **93**, 93 (2001); S. Beretta, G. Chirico, and G. Baldini, "Short-range interactions of globular proteins at high ionic strengths," *Macromolecules* **33**, 8663 (2000).

[43] L. G. Brazier, "On the flexure of thin cylindrical shells and other 'thin' sections," *Proc. Roy. Soc. London, Ser. A* **116** 104 (1927).

[44] S. Iijima, C. Brabec, A. Maiti, and J. Bernholc, "Structural flexibility of carbon nanotubes," *J. Chem. Phys.* **104**, 2089 (1996).

[45] B. I. Yakobson, C. J. Brabec, and J. Bernholc, "Nanomechanics of carbon tubes: instabilities beyond linear response," *Phys. Rev. Lett.* **76**, 2511 (1996).

[46] J. Tersoff and R. S. Ruoff, "Structural properties of a carbon-nanotube crystal," *Phys. Rev. Lett.* **73**, 676 (1994). Thave been no experimental measurements of $U$, and the theoretical literature reports a range of values (owing mainly to calculational errors). Tersoff and Ruoff give $U = 3.35$ eV nm$^{-3/2}$, while Lu [J. P. Lu, "Elastic properties of carbon nanotubes and nanoropes," *Phys. Rev. Lett.* **79**, 1297 (1997)] finds $U = 4.66$ eV nm$^{-3/2}$ (note these values differ by a factor of almost exactly $2^{1/2}$, so one value probably suffers from a calculational error rather than a difference in physical assumptions). The value given by Buldum and Lu [A. Buldum and J. P. Lu, "Atomic scale sliding and rolling of carbon nanotubes," *Phys. Rev. Lett.* **83**, 5050 (1999)] for the interaction energy of a NT with a graphite sheet implies a tube-tube value of $U = 1.34$ eV nm$^{-3/2}$. Martel, Shea, and Avouris [ref. 3] cite the paper of Tersoff and Ruoff, but the numerical values they use are inconsistent with the equations of Tersoff and Ruoff. Tersoff and Ruoff present the most verifiable details about their calculations, so we use their value of $U = 3.35$ eV nm$^{-3/2}$.

[47] A pitchfork bifurcation arises here due to the inherent symmetry in the problem; just above the onset of buckling the straight rod can acquire one of two equivalent positions that are mirror images of each other leading to the typical square-root dependence of the amplitude on the increase of the load above that at onset. A weakly nonlinear perturbation analysis yields the prefactor and hence the expression for the load as a function of the amplitude of deformation.

[48] Freely available from: http://www.susqu.edu/facstaff/b/brakke/evolver/.

[49] Simulations were done on a linear chain of 32 elements constrained to a sphere. At each timestep, the radius of the sphere was decreased by a factor of 0.01, the chain was allowed to evolve for 300 iterations to minimize its energy, and then the total strain energy in the chain was measured.

[50] SWNTs generously provided by Carbon Nanotechnologies Inc., Houston Tx.

[51] D. K. Fygenson, J. F. Marko, and A. Libchaber, "Mechanics of microtubule-based membrane extension," *Phys. Rev. Lett.* **79**, 4497 (1996).